\documentclass[useAMS,usenatbib]{mn2e}
\usepackage{graphicx}
\usepackage{url}
\usepackage[figuresright]{rotating}
\usepackage{aas_macros}
\usepackage{amssymb}
\usepackage{times}
\newcommand{\pl}{\ensuremath{\pm}}
\title[Jet-cloud collisions in NGC\,3079]{Jet-cloud collisions in the jet of the Seyfert galaxy NGC\,3079}

\author[E. Middelberg, I. Agudo, A. L. Roy, and T. P. Krichbaum]{E. Middelberg$^{1}$\thanks{E-mail:
enno.middelberg@csiro.au}, I. Agudo,$^{2,3}$, A. L. Roy$^{2}$, and T. P. Krichbaum$^{2}$\\
$^{1}$Australia Telescope National Facility, PO Box 76, Epping NSW 1710, Australia\\
$^{2}$Max-Planck-Insitut f\"ur Radioastronomie, Auf dem H\"ugel 69, 53121 Bonn, Germany\\
$^{3}$Instituto de Astrof\'{\i}sica de Andaluc\'{\i}a, CSIC, Apartado 3004, 18080 Granada, Spain}
\begin{document}

\date{}

\pagerange{\pageref{firstpage}--\pageref{lastpage}} \pubyear{2006}

\maketitle

\label{firstpage}

\begin{abstract}
We report the results from a six-year, multi-epoch very long baseline
interferomertry monitoring of the Seyfert galaxy NGC\,3079. We have
observed NGC\,3079 during eight epochs between 1999 and 2005
predominantly at 5\,GHz, but covering the frequency range of 1.7\,GHz
to 22\,GHz. Using our data and observations going back to 1985, we
find that the separation of two of the three visible nuclear radio
components underwent two decelerations. At the time of these
decelerations, the flux density of one of the components increased by
factors of five and two, respectively. We interpret these events as a
radio jet component undergoing compression, possibly as a result of a
collision with ISM material. This interpretation strongly supports the
existence of jets surrounded by a clumpy medium of dense clouds within
the first few parsecs from the central engine in NGC\,3079. Moreover,
based on recently published simulations of jet interactions with
clumpy media, this scenario is able to explain the nature of two
additional regions of ageing synchrotron material detected at the
lower frequencies as by-products of such interactions, and also the
origin of the kpc-scale super bubble observed in NGC\,3079 as the
result of the spread of the momentum of the jets impeded from
propagating freely. The generalization of this scenario provides an
explanation why jets in Seyfert galaxies are not able to propagate to
scales of kpc as do jets in radio-loud AGN.

\end{abstract}


\section{Introduction}

Active galactic nuclei (AGN) are conventionally divided into
radio-loud and radio-quiet objects, depending on their ratio of
radio-to-optical luminosity. Although radio-quiet objects such as
Seyfert galaxies have luminosities of up to 1000 times less than
radio-loud objects, they mostly exhibit high-brightness temperature
(non-thermal) components (\citealt{Middelberg2004}), indicating that
the radio emission is synchrotron emission. However, it is not clear
why the generation of radio emission in Seyfert galaxies is different
from that in radio-loud objects such as quasars and radio galaxies. In
particular, Seyferts mostly do not exhibit long and confined jets as
are typically observed in radio-loud objects. Only in few radio-quiet
objects have jet-like features been observed, prominent examples are
NGC\,4258 (\citealt{Miyoshi1995}), NGC\,4151 (\citealt{Mundell2003}),
NGC\,3079 (\citealt{Irwin1988,Trotter1998,Kondratko2005}), and
NGC\,1068 (\citealt{Roy1998}),

NGC\,3079 is a nearby (15.0\,Mpc, \citealt{Vaucouleurs1991}, ${\rm
H_0=75\,km\,s^{-1}\,Mpc^{-1}}$, $1\,{\rm mas\,yr^{-1}=0.24}\,c$) LINER
(\citealt{Heckman1980}) or Seyfert~2 (\citealt{Sosa-Brito2001}) galaxy
in an edge-on, dusty spiral. In the radio, pilot VLBI observations by
\cite{Irwin1988} resolved the core into two strong components, $A$ and
$B$, separated by 20.2\,mas. VLBI observations carried out by
\cite{Trotter1998}, \cite{Sawada-Satoh2000}, \cite{Kondratko2005} and
\cite{Middelberg2005c} revealed that the separation of $A$ and $B$ was
increasing at a speed between $0.11\,c$ and $0.16\,c$ between 1985 and
2000. On the axis connecting $A$ and $B$, indications of another
component called $C$ were already found by
\cite{Irwin1988}. With very sensitive observations it has recently
been possible to detect a structure that covers about half the
projected distance of 2\,pc between $A$ and $B$
(\citealt{Kondratko2005}), indicating a jet-like feature. NGC\,3079
therefore appears to be a promising candidate source to investigate
the nature of radio emission in Seyferts.

NGC\,3079 also contains one of the brightest known extragalctic water
masers ($\nu_{\rm rest}=22.23508\,{\rm GHz}$), which has been subject
to investigation for more than a decade (e.g.,
\citealt{Haschick1985,Haschick1990,Trotter1998,Kondratko2005}). The
distribution of the masers and their velocity dispersion indicate a
thick molecular disk rotating around a massive object, presumably a
black hole.

We have observed the radio emission in NGC\,3079 with the VLBA at
5\,GHz, 15\,GHz, and 22\,GHz, using phase referencing in most epochs,
to monitor precisely its structural evolution. Single epoch images at
1.7\,GHz and 2.3\,GHz were obtained to probe a wider spectral range.

\section{Observations}

The data presented here were obtained with the VLBA over eight epochs
between 1999 and 2005, partly supplemented by other telescopes. The
sensitivity, $(u,v)$ coverage, and resolution varied due to the wide
range in frequency. Also, phase referencing was used in the later
observations, but not in the 1999/2000 observations. A summary of the
observations is presented in Table~\ref{tab:arrays}.

At 1.7\,GHz and 2.3\,GHz, NGC\,3079 was observed during a single epoch
only, on 22 September 2002, using phase referencing to the nearby
calibrator 4C\,+55.17. The Effelsberg telescope was included in the
1.7\,GHz observations only, to increase the resolution at this low
frequency.

We have observed NGC\,3079 at 5\,GHz during eight epochs between
November 1999 and August 2005 with the VLBA, in part supplemented by
the Effelsberg telescope and a single VLA antenna. The observations
carried out in 2002-2005 were phase-referenced to the nearby source
4C\,+55.17. We have also obtained a 5\,GHz VLBI image from a
16-station global VLBI snapshot observation performed in 1994 (project
GK011), but were only able to reliably measure the separation of $A$
and $B$ from this observation.

During three epochs in 1999-2000, we have observed NGC\,3079 at
15\,GHz with the VLBA, the Effelsberg telescope and a single VLA
antenna.  The observations yielded only few detections on some
baselines shorter than 100\,M$\lambda$ during the November 1999 and
April 2000 epochs, but in November 2000, the source was detected on
most baselines shorter than 120\,M$\lambda$.

During the three epochs in 2004-2005, we observed the water maser
emission in NGC\,3079 at $\nu_{\rm rest}=22.23508\,{\rm GHz}$ and the
22\,GHz continuum emission. The frequency resolution was 31.25\,kHz,
corresponding to a velocity resolution of $0.42\,{\rm
km\,s^{-1}}$. The continuum emission was phase-referenced to the
brightest maser feature.

\begin{table*}
\centering
\begin{tabular}{l|lllll}
\hline
Date            & $\nu$         & suppl. Antennas       & $\Delta t$    & $\Delta \nu$  & Phase-ref\\
                &  GHz          &                       & min           & MHz           & \\
\hline
1994-02-28      & 5.0           & Eb/Y/On/Me/No         & $\sim$10      & 56            & n \\
1999-11-20      & 1.7, 5.0, 15  & Y1/Eb,Y1/Eb,Y1        & 108, 144, 144 & 32            & n \\
2000-03-06      & 1.7, 5.0, 15  & Y1/Eb,Y1/Eb,Y1        & 108, 144, 136 & 32            & n \\
2000-11-30      & 1.7, 5.0, 15  & Y1/Eb,Y1/Eb,Y1        & 108, 144, 144 & 32            & n \\
2002-09-22      & 1.7, 2.3, 5.0 & Eb                    & 114, 107, 112 & 64            & y \\
2004-06-14      & 5.0, 22       &                       & 199, 94       & 64            & y \\
2004-11-03      & 5.0, 22       &                       & 194, 82       & 64            & y \\
2005-04-04      & 5.0, 22       &                       & 195, 107      & 64            & y \\
2005-08-18      & 5.0, 22       &                       & 192, 107      & 64            & y \\
\hline
\end{tabular}
\caption{Summary of the observations. We give the observing dates, the
frequencies observed, antennas supplementing the VLBA (Eb -
Effelsberg, Y1 - one VLA antenna, Y - phased VLA, Me - Medicina, On -
Onsala 85, No - Noto), the integration times, the bandwidths, and whether
phase referencing was used or not. 2-bit sampling was used except
1994, when 1-bit sampling was used.}
\label{tab:arrays}
\end{table*}

\subsection{Calibration}

The data were calibrated using standard procedures implemented in the
Astronomical Image Processing System, AIPS. The visibility amplitudes
were calibrated using noise-adding radiometry and established antenna
gains, and amplitude errors of the order of 5\,\% arising from sampler
threshold variations were corrected using the autocorrelation
data. Phase rotation due to parallactic angle changes was corrected,
and instrumental delays were calibrated using fringe-fitting on a
short observation of a strong source. For phase calibration of the
observations carried out in 1999-2000, we fringe-fitted the data using
a point source model, with solution intervals of around 4\,min,
yielding detections on baselines up to 100\,M$\lambda$ at both 5\,GHz
and 15\,GHz. At 22\,GHz, the phase calibration was carried out using
six frequency channels with a total bandwidth of 187.5\,kHz for
self-calibration on the strongest water maser feature, at a velocity
of 956\,km\,s$^{-1}$. The maser was detected on most baselines, and
the phase solutions were then applied to all frequency channels.

\subsection{Eearth orientation parameter errors and phase calibrator
position changes}

The observations in 2004 and 2005 were correlated using predicted
rather than measured earth orientation parameters, a situation arising
from a software upgrade at the VLBA correlator. Furthermore, the
position of the phase calibrator 4C\,+55.17 used for phase referencing
of the 5\,GHz observations in the two 2005 epochs differed from the
position used in earlier epochs. This happened because calibrator
positions in the program used to schedule VLBA observations, SCHED,
were taken from a data base at the Array Operations Center at the time
of observing, rather than from the schedule, and hence were subject to
unforseeable changes. However, the effects of both problems are
predictable and precisely known, and were easily rectified.

We retrieved the final earth orientation parameters from the USNO
solution at
\url{http://gemini.gsfc.nasa.gov/solve_save/usno_finals.erp} and 
applied the corrections to the data using the AIPS task VLBAEOP. This
process has been well tested and the residual phase errors after
correction are of the order of $<5^\circ$ at 5\,GHz
(\citealt{Walker2005}). The position of the calibrator in the
2002-2004 observations, 4C\,+55.17, was modified using the AIPS task
CLCOR. The right ascension was decreased by 3.375\,mas and the
declination was decreased by 0.792\,mas, to agree with the presumably
more accurate position used in the 2005 observations,
RA=09:57:38.184971, Dec=55:22:57.76924 (J2000).

\subsection{Image analysis}
\label{sec:analysis}

The relative positions and flux densities of the 5\,GHz and 15\,GHz
observations in 1994 and 1999-2000 were measured from self-calibrated
images (because no phase referencing was carried out), tapered to a
resolution of $3\times3\,{\rm mas}^2$ as a compromise between $(u,v)$
coverage and sensitivity. Absolute positions were not measured in
these observations.

The absolute positions of the 5\,GHz and 22\,GHz observations in
2002-2005 were measured from purely phase-referenced images, with no
self-calibration steps carried out, and tapered to $3\times3\,{\rm
mas}^2$. The flux densities at 5\,GHz were measured from
phase-referenced, self-calibrated, and tapered images, to remove
residual phase errors and improve the signal-to-noise ratio. Flux
densities at 22\,GHz were measured from purely phase-referenced images
because self-calibration turned out to be marginal. In
self-calibration, the resulting flux densities of components $A$ and
$B$ were found to vary by as much as a factor of three when the
calibration strategy was changed only very little and our models did
not reliably converge. We therefore consider the 22\,GHz flux
densities to be very unreliable and do not use them in our analysis.

We have measured the positions of components $A$, $B$, and $E$ and
their peak flux densities by fitting a paraboloid to a square of three
by three pixels, using the location of the brightest pixel as a
starting point.  Exploring five images from a reasonable imaging
parameter space revealed a position uncertainty of 0.15\,mas
($1\,\sigma$) for the 5\,GHz and 15\,GHz images, and of 0.1\,mas for
the 22\,GHz images. Hence separations, which include two position
measurements, have errors of $\sqrt{2}$ times the position errors. The
errors of our 1994 observations are larger because of the very sparse
$(u,v)$ coverage. 

Integrated flux densities were measured by integrating over the source
region. We have estimated the flux-density errors as follows: peak
flux densities have a 5\,\% (10\,\%) scaling error at frequencies at
and below 5\,GHz (15\,GHz) plus the image rms added in
quadrature. Integrated flux densities have the same scaling errors,
plus the image rms in quadrature, plus an error arising from the
integration over the source region, which was derived experimentally,
added in quadrature. For components $A$ and $F$, the integration error
alone amounts to 1\,mJy because they blend with $C$ or are poorly
defined, and to 0.3\,mJy for components $B$ and $E$.

\begin{figure}
\centering
\includegraphics[width=\linewidth]{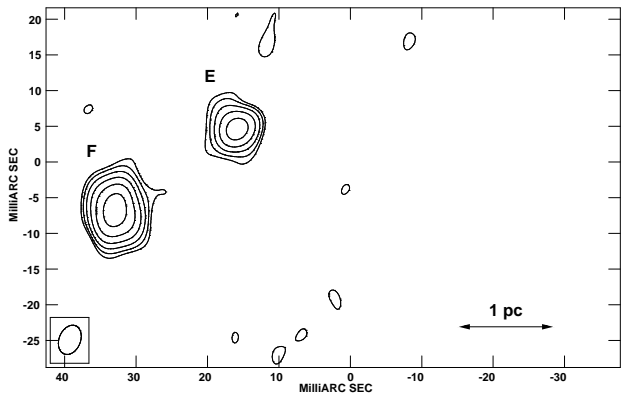}
\caption{1.7\,GHz, phase-referenced and self-calibrated image of
NGC\,3079, observed on 22 September 2002. Axes are relative to RA
10h01m57.7906s Dec $55^\circ40'47.1788''$ (J2000), and
self-calibration produced no measureable shift in the coordinates from
phase referencing. Contours are drawn at $0.18\,{\rm
mJy\,beam^{-1}}\times2^N~(N=0,1,2,...)$, and the beam size is
$4.26\times2.99\,{\rm mas}^2$. Only components $E$ and $F$ have been
detected, and a wider region is shown to demonstrate that $A$, $B$,
and $C$ are not visible.  }
\label{fig:3079_L}
\end{figure}

\begin{figure}
\centering
\includegraphics[width=\linewidth]{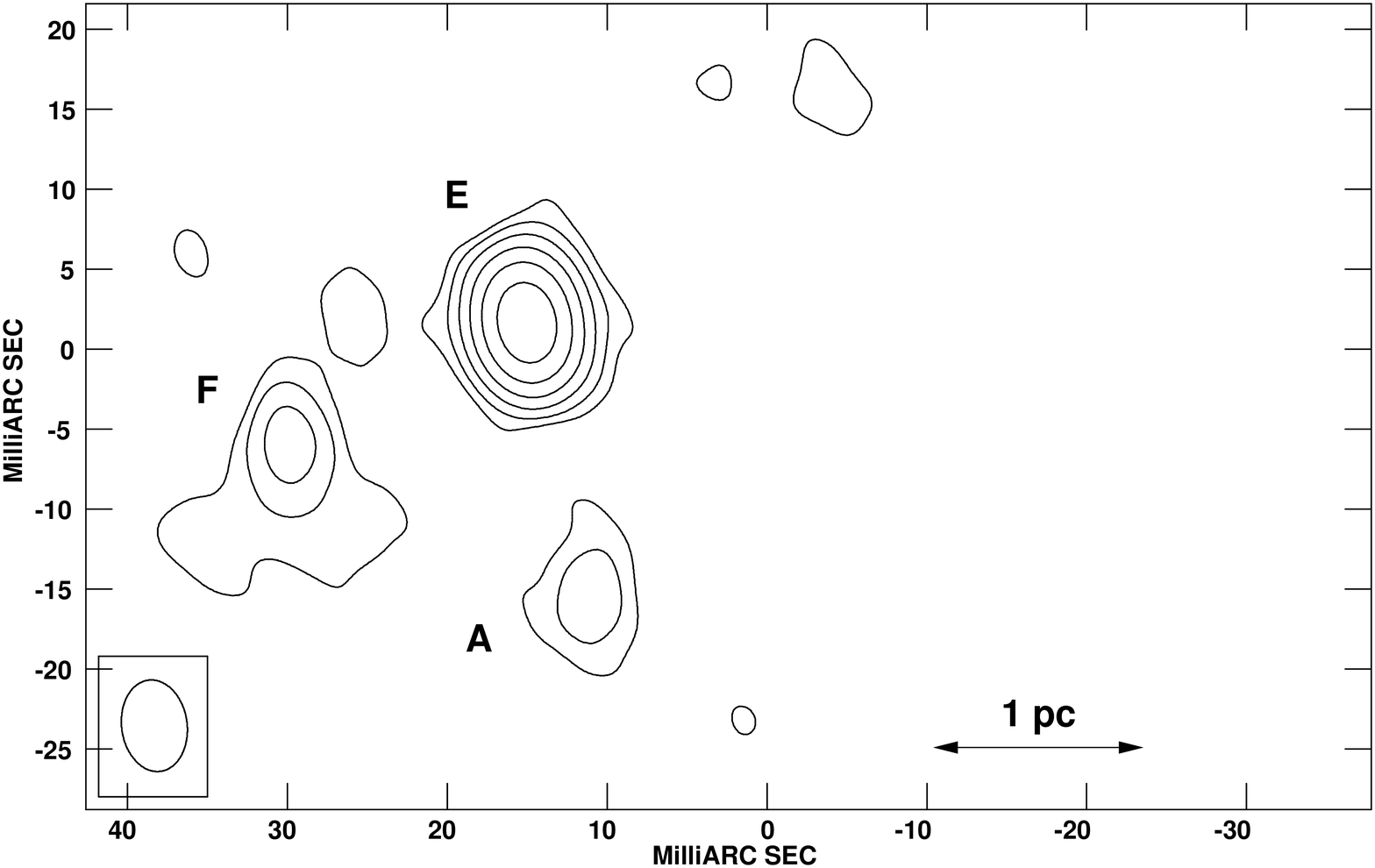}
\caption{2.3\,GHz, phase-referenced and self-calibrated image of
NGC\,3079, observed on 22 September 2002.  Axes are relative to RA
10h01m57.7906s Dec $55^\circ40'47.1788''$ (J2000), and
self-calibration produced no measureable shift in the coordinates from
phase referencing. Contours are at $0.39\,{\rm
mJy\,beam^{-1}}\times2^N~(N=0,1,2,...)$, and the beam size is
$5.75\times4.10\,{\rm mas}^2$. Components $E$ and $F$ are weaker than
at 1.7\,GHz, and $A$ is weak, but detected.}
\label{fig:3079_S}
\end{figure}

\begin{figure}
\centering
\includegraphics[width=\linewidth]{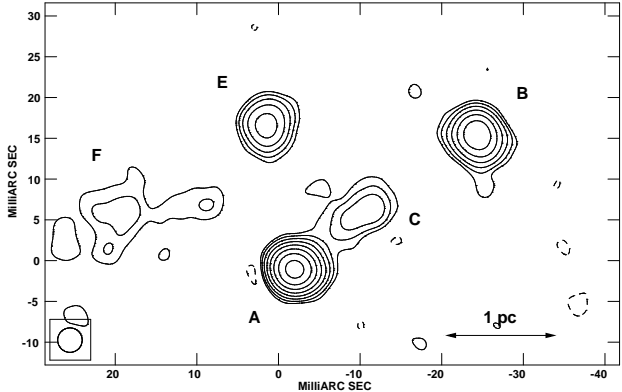}
\caption{Example of a 5\,GHz phase-referenced and self-calibrated image
of NGC\,3079, observed on 18 August 2005. At this frequency, all
components can be seen. Components $A$, $B$, and $E$ are still strong,
the jet-like feature $C$ between $A$ and $B$ is clearly detected, and
$F$ is resolved and faint, but detected. Axes are relative to RA
10h01m57.7906s Dec $55^\circ40'47.1788''$ (J2000), and
self-calibration produced no measurable shift in the coordinates from
phase referencing. Contours are drawn at $0.24\,{\rm
mJy\,beam^{-1}}\times2^N~(N=0,1,2,...)$. The data have been tapered to
a resolution of $3\times3\,{\rm mas}^2$.}
\label{fig:3079_C}
\end{figure}

\begin{figure}
\centering
\includegraphics[width=\linewidth]{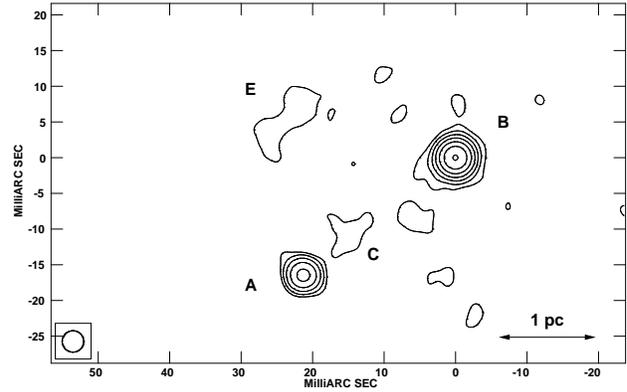}
\caption{Example of a 15\,GHz image of NGC\,3079, observed on 30
November 2000. Components $A$ and $B$ are clearly detected, but there
are only hints of components $C$ and $E$. Absolute positions were not
measured (phase referencing was not used), contours are at $0.81\,{\rm
mJy\,beam^{-1}}\times2^N~(N=0,1,2,...)$. The data have been tapered to
a resolution of $3\times3\,{\rm mas}^2$.}
\label{fig:3079_U}
\end{figure}

\begin{figure}
\centering
\includegraphics[width=\linewidth]{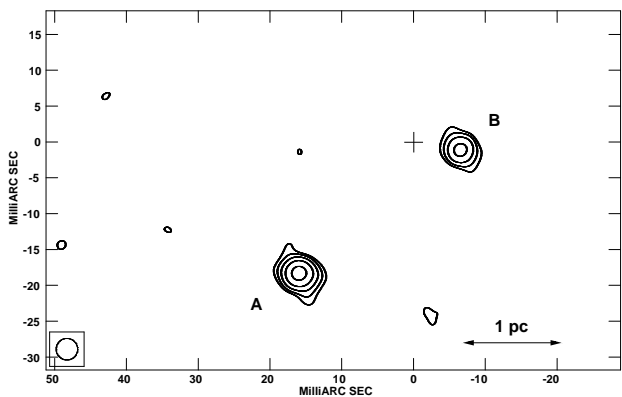}
\caption{Example of a 22\,GHz continuum image of NGC\,3079, observed on 18 Aug
2005. Only components $A$ and $B$ have been detected, and then with
relatively low SNR. Phase referencing to the brightest maser did not
yield absolute positions at this frequency, contours are at $6.3\,{\rm
mJy\,beam^{-1}}\times2^N~(N=0,1,2,...)$. The data have been tapered to
a resolution of $3\times3\,{\rm mas}^2$, which caused a large increase
of the image noise. The cross indicates the position of the brightest
maser feature.}
\label{fig:3079_K}
\end{figure}

For illustration, we show in Figures~\ref{fig:3079_L}-\ref{fig:3079_K}
1.7\,GHz and 2.3\,GHz images from the 22 September 2002 epoch, 5\,GHz
and 22\,GHz images from the 18 August 2005 epoch, and a 15\,GHz image
from the 30 November 2000 epoch. Details can be found in the image
captions.

{\bf Caption to Table~2} Component and image
parameters. Note that the 22\,GHz flux densities are not reliable
because self-calibration was marginal, see the discussion in
Section~\ref{sec:analysis}.

\newpage

\begin{sidewaystable*}
\begin{tabular}{l|cccccccc|cccc}
\hline
Date            & \multicolumn{8}{c|}{Component parameters}                                            & \multicolumn{4}{c}{Image parameters}\\
		& \multicolumn{4}{c}{Peak flux density} & \multicolumn{4}{c|}{Int. flux density} & Image rms & $b_{\rm maj}$ & $b_{\rm min}$ & PA\\
		& \multicolumn{4}{c}{mJy\,beam$^{-1}$}  & \multicolumn{4}{c|}{mJy}               & mJy\,beam$^{-1}$  & mas           & mas           & $^\circ$\\
                & $A$         & $B$         & $E$         &  $F$        & $A$         & $B$         & $E$         & $F$    &        \\
\hline
{\it Observations at 1.7\,GHz}&&&&&&&&&&&\\
2002-09-22      &             &             & 4.2\pl0.2   & 9.7\pl0.5   &             &             & 5.4\pl0.6	  & 15.3\pl1.3& 0.06   & 5.1 & 3.5 & -22\\
\hline
{\it Observations at 2.3\,GHz}&&&&&&&&&&&\\
2002-09-22      & 1.2\pl0.1   &             & 19.8\pl1.0  & 2.1\pl0.2   & 1.9\pl1.0  &             & 21.0\pl1.2  & 7.2\pl0.7 & 0.13   & 6.0 & 4.3 & 12\\
\hline
{\it Observations at 5\,GHz}&&&&&&&&&&&\\
1999-11-20      & 24.8\pl1.2  & 12.8\pl0.7  & 3.1\pl0.2   &             & 27.1\pl1.7  & 15.9\pl1.0  & 4.1\pl0.6   &           & 0.097  & 3.0 & 3.0 & 0.0\\
2000-03-06      & 23.3\pl1.2  & 13.1\pl0.7  & 3.1\pl0.2   &             & 26.0\pl1.6  & 18.7\pl1.1  & 3.2\pl0.6   &           & 0.095  & 3.0 & 3.0 & 0.0\\
2000-11-30      & 22.9\pl1.1  & 13.2\pl0.7  & 3.8\pl0.2   &             & 23.8\pl1.6  & 18.8\pl1.1  & 6.4\pl0.6   &           & 0.092  & 3.0 & 3.0 & 0.0\\
2002-09-22      & 21.1\pl1.1  & 14.0\pl0.7  & 5.4\pl0.3   &             & 26.0\pl1.6  & 22.4\pl1.3  & 7.9\pl0.7   & 6.9\pl1.1 & 0.089  & 3.0 & 3.0 & 0.0\\
2004-06-14      & 41.7\pl2.1  & 16.6\pl0.8  & 5.4\pl0.3   &             & 47.0\pl2.6  & 21.8\pl1.2  & 8.9\pl0.7   &           & 0.085  & 3.0 & 3.0 & 0.0\\
2004-11-03      & 41.7\pl2.1  & 15.8\pl0.8  & 5.3\pl0.3   &             & 47.5\pl2.6  & 21.2\pl1.2  & 7.9\pl0.7   &           & 0.084  & 3.0 & 3.0 & 0.0\\
2005-04-04      & 40.5\pl2.0  & 15.2\pl0.8  & 5.7\pl0.3   &             & 46.1\pl2.5  & 21.7\pl1.2  & 9.1\pl0.7   &           & 0.067  & 3.0 & 3.0 & 0.0\\
2005-08-18      & 42.8\pl2.1  & 14.6\pl0.7  & 5.6\pl0.3   &             & 50.1\pl2.7  & 21.1\pl1.2  & 9.2\pl0.7   &           & 0.080  & 3.0 & 3.0 & 0.0\\
\hline
{\it Observations at 15\,GHz}&&&&&&&&&&&\\
1999-11-20      & 6.7 \pl0.8  & 46.2\pl4.6  &       	  &       	& 8.1 \pl1.3  & 50.0\pl5.0  &       &        & 0.39   & 3.0 & 3.0 & 0.0\\
2000-03-06      & 7.6 \pl0.9  & 53.6\pl5.4  &       	  &       	& 9.5 \pl1.5  & 55.4\pl5.6  &       &        & 0.46   & 3.0 & 3.0 & 0.0\\
2000-11-30      & 16.2\pl1.6  & 53.7\pl5.4  &       	  &       	& 17.6\pl2.0  & 57.8\pl5.8  &       &        & 0.27   & 3.0 & 3.0 & 0.0\\
\hline
{\it Observations at 22\,GHz}&&&&&&&&&&&\\
2004-11-03      & (28.6)      & (81.9)      &       	  &       	& (28.6)      & (80.6)      &       &        & 0.99   & 3.0 & 3.0 & 0.0\\
2005-04-04      & (49.0)      & (70.1)      &       	  &       	& (54.1)      & (69.8)      &       &        & 0.84   & 3.0 & 3.0 & 0.0\\
2005-08-18      & (132.8)     & (62.5)      &       	  &       	& (147.1)     & (67.1)      &       &        & 2.09   & 3.0 & 3.0 & 0.0\\
\hline
\end{tabular}
\label{tab:obs}
\end{sidewaystable*}

\addtocounter{table}{1}

\begin{table*}
\centering
\begin{tabular}{lcccc}
\hline
Date		& r(B-A) & PA(A)    & r(B-E) & PA(E)\\
                & mas    & $^\circ$ & mas    & $^\circ$\\
\hline
{\it Observations at 5\,GHz}\\
1994-02-28      & 24.03$\pm$0.42  & 126.6$\pm$1.0 \\
1999-11-20	& 26.22$\pm$0.21  & 126.3$\pm$0.5  & 26.26$\pm$0.21  & 86.7$\pm$0.5 \\
2000-03-06	& 26.41$\pm$0.21  & 126.2$\pm$0.5  & 26.17$\pm$0.21  & 86.9$\pm$0.5 \\
2000-11-30	& 26.53$\pm$0.21  & 126.6$\pm$0.5  & 25.91$\pm$0.21  & 87.4$\pm$0.5 \\
2002-09-22	& 27.20$\pm$0.21  & 127.4$\pm$0.5  & 25.77$\pm$0.21  & 88.5$\pm$0.5 \\
2004-06-14	& 27.77$\pm$0.21  & 126.9$\pm$0.5  & 25.89$\pm$0.21  & 88.7$\pm$0.5 \\
2004-11-03	& 27.79$\pm$0.21  & 126.8$\pm$0.5  & 26.13$\pm$0.21  & 88.6$\pm$0.5 \\
2005-04-04	& 27.69$\pm$0.21  & 126.8$\pm$0.5  & 25.79$\pm$0.21  & 87.9$\pm$0.5 \\
2005-08-18	& 27.85$\pm$0.21  & 126.5$\pm$0.5  & 25.79$\pm$0.21  & 88.5$\pm$0.5 \\
{\it Observations at 15\,GHz}\\
1999-11-20	& 26.82$\pm$0.21  & 128.0$\pm$0.5 \\
2000-03-06	& 26.76$\pm$0.21  & 127.5$\pm$0.5 \\
2000-11-30	& 26.77$\pm$0.21  & 127.2$\pm$0.5 \\
\hline
{\it Observations at 22\,GHz}\\
2004-11-03      & 28.24$\pm$0.14  & 127.0$\pm$0.3 \\
2005-04-04      & 28.39$\pm$0.14  & 127.2$\pm$0.3 \\
2005-08-18      & 28.50$\pm$0.14  & 126.8$\pm$0.3 \\
\hline
\end{tabular}
\caption{Distances (r) and position angles (PA) 
of components A and E with respect to B.}
\label{tab:pos}
\end{table*}

\section{Results}

\subsection{The maser emission}

\begin{figure}
\centering
\includegraphics[width=\linewidth]{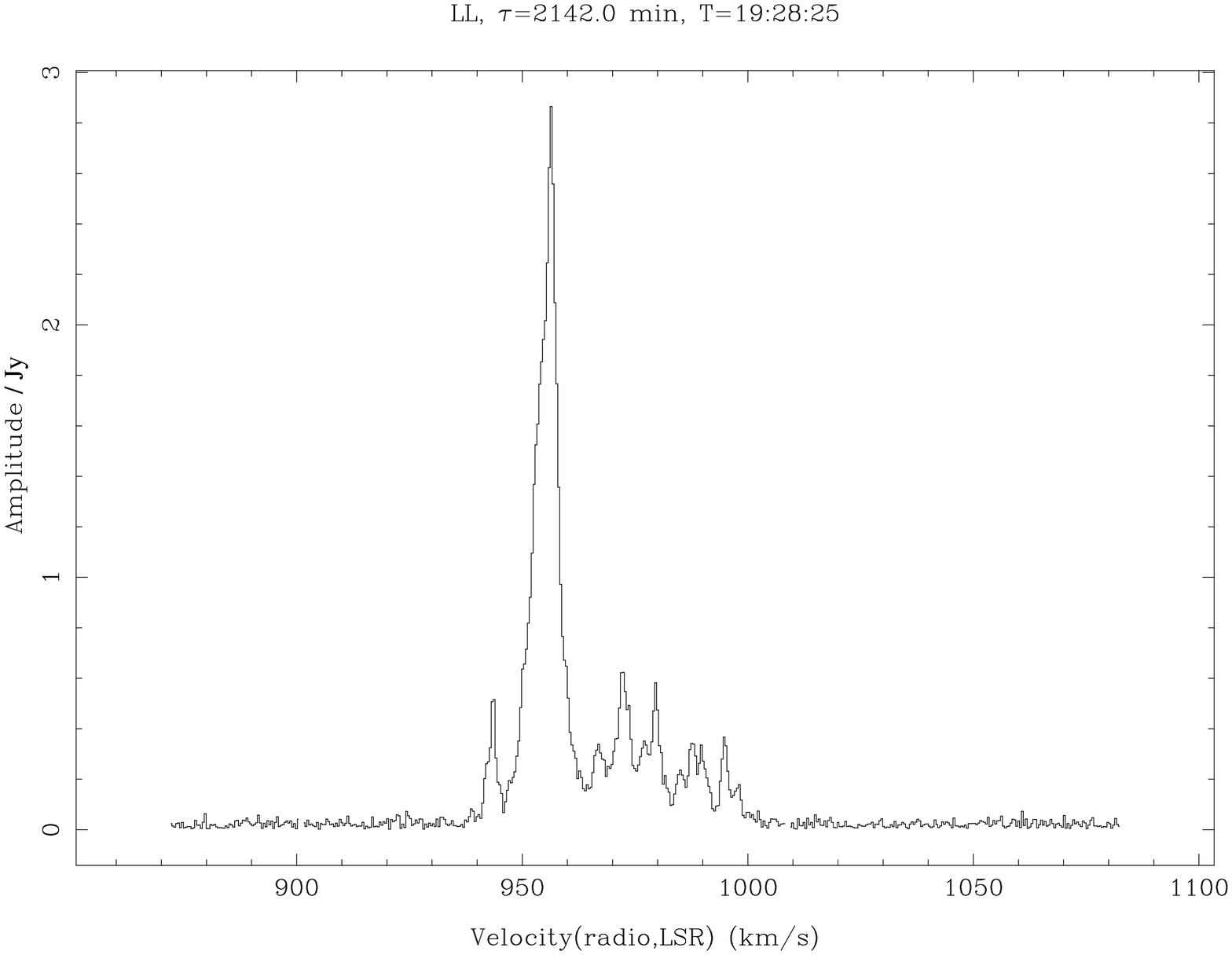}  
\caption{A spectrum of the water maser emission in NGC\,3079, derived
from averaging the cross-power spectra of all baselines. No lines were
detected outside the velocity range shown.}
\label{fig:spectrum}
\end{figure}

The purpose of observing the maser emission in this project was to
phase reference the radio continuum emission to a quasi-stationary
point (cf. the next section). In our data, the $1\,\sigma$ thermal
noise per channel is $12\,{\rm mJy\,beam^{-1}}$, only allowing us
to detect the brightest features. We therefore do not attempt an
astrophysical modelling of the maser emission. A spectrum showing all
detected lines is shown in Figure~\ref{fig:spectrum}.

\begin{figure}
\centering
\includegraphics[width=\linewidth]{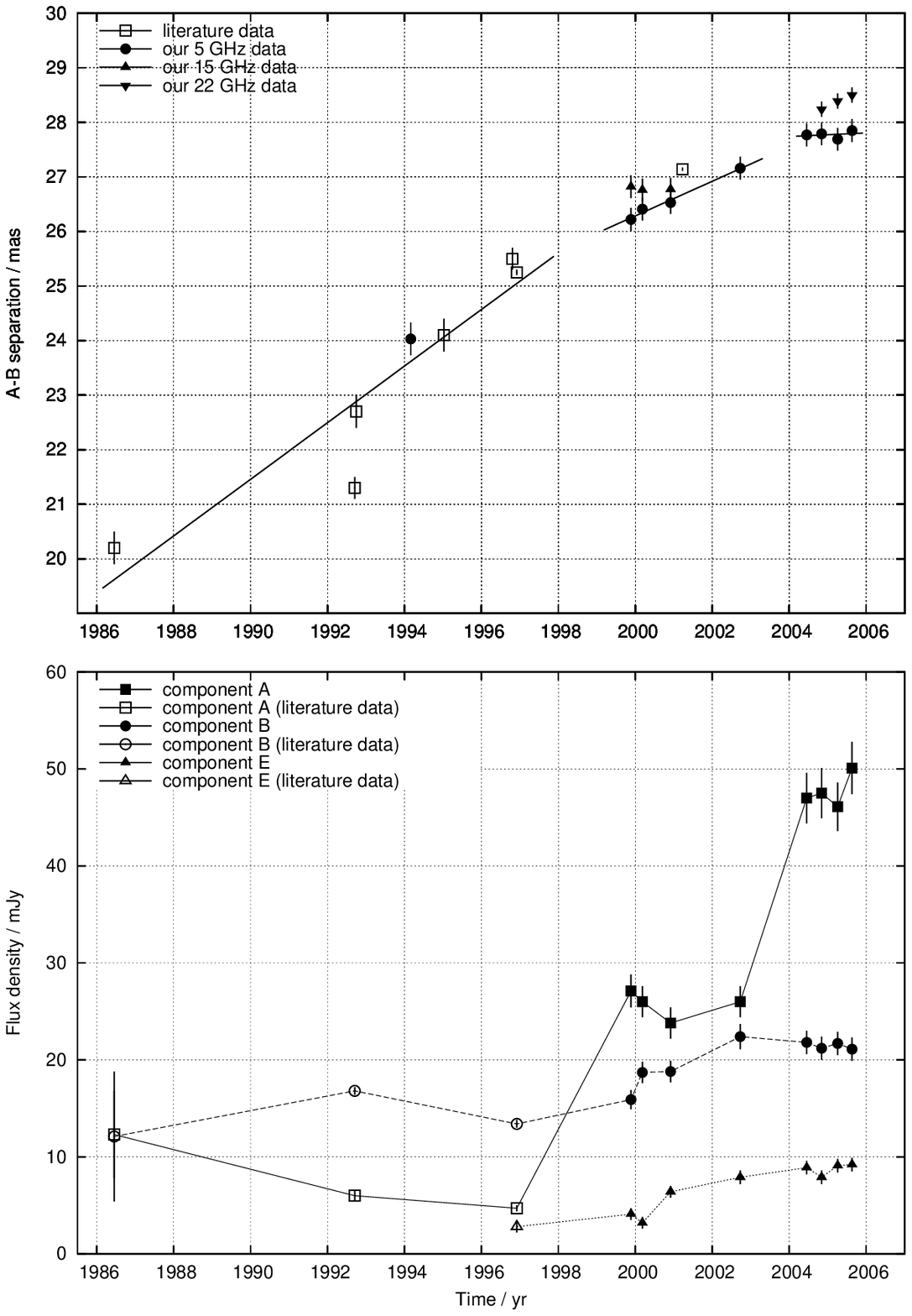}  
\caption{{\it Upper panel:} The evolution of the separation of $A$ and $B$,
using data from the literatrure and our measurements. Literature data
are denoted with open symbols and have been taken from
\citet{Irwin1988} (epoch 1986.455, 5\,GHz), \citet{Trotter1998} (epochs
1992.706, 5\,GHz, and 1992.745, 8\,GHz), \citet{Sawada-Satoh2000}
(epoch 1996.802, 8\,GHz), and \citet{Kondratko2005} (epochs 1996.919,
5\,GHz, and 2001.225, 22\,GHz) {\it Lower panel:} The 5\,GHz flux
densities of components $A$, $B$, and $E$. Only 5\,GHz literature
data, denoted with open symbols, were incorporated in this diagram and
were taken from the same publications as for the panel
above. Measurements of $A$, $B$, and $E$ have been connected with
solid, dashed, and dotted lines to help guide the eye.}
\label{fig:sep+fluxes}
\end{figure}

\subsection{Position of $B$ with respect to the maser emission}

One of the aims of the 22\,GHz observations was to measure the
positions of $A$ and $B$ with respect to the water maser emission. We
found that the separation of the brightest maser feature at
956\,km\,s$^{-1}$ from the continuum component $B$ was $6.7\,{\rm
mas}\pm0.14\,{\rm mas}$, with a position angle of 82$^\circ$. This is
in agreement with measurements by \cite{Trotter1998} ($6.7\,{\rm
mas}\pm0.1\,{\rm mas}$ in PA $81^\circ$), \cite{Sawada-Satoh2002}
($\approx6.7\,{\rm mas}$ in PA $81^\circ$), and
\cite{Kondratko2005} ($6.74\,{\rm mas}\pm0.13\,{\rm mas}$ in PA
$80.75^\circ\pm1.04^\circ$), which do not reveal any conclusive motion
between January 1995 and March 2001. This means that component $B$ has
been stationary with respect to the maser emission for a decade. If
the masers trace a molecular disk around the central massive black
hole (\citealt{Kondratko2005, Trotter1998}), and their velocity
dispersion is 400\,km\,s$^{-1}$, or 0.0013\,$c$, then masers moving
parallel to the plane of the sky could have moved by about
0.013\,l.y., or 0.004\,pc in 10\,yr. This corresponds to 0.06\,mas,
which would not have been noticed. Hence $B$ can be regarded as
stationary with respect to the central black hole powering the AGN,
and measuring proper motions relative to $B$ is therefore equivalent
to measuring them relative to the black hole.


\subsection{Separation of $A$ and $B$}
\label{sec:A-B}

VLBI observations of NGC\,3079 have shown early on that the separation
of $A$ and $B$ is increasing (\citealt{Middelberg2005c},
\citealt{Kondratko2005}). We have plotted the previously published
measurements of the separation of $A$ and $B$ in
Fig.~\ref{fig:sep+fluxes}, together with our new measurements.

The pre-1998 data show a steady expansion of the source, with the
notable exception of the 5\,GHz measurement by \cite{Trotter1998}. The
data taken between 1999 and 2003 already indicate a slightly slower
expansion rate, and the 2004-2005 5\,GHz data clearly show that the
expansion has come to a halt. We note that the separation of $A$ and
$B$ is consistently larger at higher frequencies, and we therefore
restrict our analysis of the proper motions to 5\,GHz whenever
possible.

The separations measured at higher frequencies do not follow the same
trend as the 5\,GHz measurements. The separations at both 15\,GHz and
22\,GHz are larger than that at 5\,GHz; the 15\,GHz data from
1999-2000 imply a larger separation and do not seem to confirm the
expansion seen at 5\,GHz; the 22\,GHz data even seem to indicate that
the expansion is continuing. However, this apparent disagreement on
the evolution of the separation of $A$ and $B$ at low (5\,GHz) and
high (15\,GHz and 22\,GHz) frequencies can be explained as a
consequence of optical depth effects causing a frequency-dependent
core shift in either or both components. We postulate that at 5\,GHz
we see only an outer layer of the component(s) where the general
kinematic properties of the whole actual emission region could be
defined. At 15\,GHz and 22\,GHz, the optical depth of the ambient
medium allows us to see predominantly into a denser emission region
with slightly different kinematic properties.

\begin{figure}
\centering
\includegraphics[width=0.5\linewidth, angle=270]{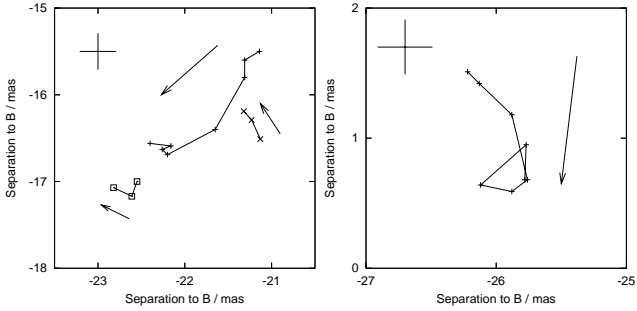}  
\caption{The motions of components $A$ (left panel) and $E$ (right
panel) with respect to component $B$. Pluses represent measurements at
5\,GHz, crosses those at 15\,GHz, and squares those at 22\,GHz. Arrows
indicate the direction of motion of a component. The large crosses in
the upper left corners indicate the errors of the measurements.}
\label{fig:xy}
\end{figure}

We have fitted linear functions to all separations observed in
1985-1996, to our 5\,GHz observations from 1999-2002, and to our
5\,GHz observations from 2004-2005. The results are shown in
Table~\ref{tab:fits} and are drawn as solid lines in
Figure~\ref{fig:sep+fluxes}. We illustrate the motion on the sky of
$A$ and $E$ with respect to $B$ in Figure~\ref{fig:xy}.

\begin{table}
\begin{center}
\begin{tabular}{lcccc}
\hline
\hline
Period    & $v_{\rm ang}$ & $\Delta v_{ang}$ & $\beta$ & $\Delta\beta$\\
          & mas\,yr$^{-1}$ & mas\,yr$^{-1}$\\
\hline
1985-1996 & 0.52 & 0.10 & 0.12 & 0.02\\
1999-2002 & 0.32 & 0.03 & 0.08 & 0.01\\
2004-2005 & 0.03 & 0.09 & 0.01 & 0.02\\
\hline
\end{tabular}
\caption{Results of the linear fits to the separation of $A$ with respect to
$B$ as a function of time. $v_{\rm ang}$ and $\Delta v_{ang}$ are the
apparent angular speed and its associated error, and $\beta$ and
$\Delta\beta$ are the corresponding values of $\beta=v/c$, assuming
that the motion is in the plane of the sky at a distance of 15\,Mpc.}
\label{tab:fits}
\end{center}
\end{table}

\subsection{Motion of component $E$}

Contrary to the separation of $A$ and $B$, which appeared to increase
steadily over more than a decade, the motion of $E$ is more
erratic. Its separation from component $B$ is plotted in the right
panel of Figure~\ref{fig:xy} and an arrow indicates the direction of
motion. During our observations in 1999 and 2000, $E$ appeared to be
moving inwards, approximately in direction of the dynamical
centre. This relatively clear trend stopped between November 2000 and
our follow-up observations in September 2002. $E$ has since described
a figure resembling a circle, and in our latest measurements in August
2005 reached a point where it had been in November 2000.

\subsection{Flux densities of $A$, $B$, and $E$}

The evolution of the 5\,GHz integrated flux densities of $A$, $B$, and
$E$ is shown in Fig.~\ref{fig:sep+fluxes}. The flux density of $A$
exhibits two sudden increases, the first between the 1996 observations
of \cite{Kondratko2005} and our observations in November 1999, and the
other between our September 2002 and June 2004 observations. The mean
of the pre-1999 flux densities of A is $(7.7\pm4.1)\,{\rm mJy}$, of
our measurements between 1999 and 2002 is $(26.0\pm1.6)\,{\rm mJy}$,
and of our post-2004 measurements is $(48.5\pm1.9)\,{\rm mJy}$. More
importantly, these increases in flux density are coincident with the
decelerations of component $A$ as described in
Section~\ref{sec:A-B}. The flux density of $B$ shows no conclusive
evolution over the years, whereas that of component $E$ appears to be
increasing slowly. The shape and internal structure of $E$ evolved
significantly over the years, and part of its erratic flux density
evolution may be ascribed to this observation.


We note that in our last 5\,GHz epoch, the brightness temperatures of
$A$, $B$, and $E$ were $39\times10^7\,{\rm K}$, $17\times10^7\,{\rm
K}$, and $7\times10^7\,{\rm K}$, indicating that the emission is
synchrotron emission from AGN or supernovae. Brightness temperatures
of HII regions generally reach only of the order of $10^4\,{\rm K}$
(e.g., \citealt{Condon1991}), and are extended with sizes of
$>>0.1^{\prime\prime}$ ($>>7.2\,{\rm pc}$ in NGC\,3079).

\subsection{Spectral indices}

\renewcommand{\arraystretch}{1.25}
\begin{table*}
\scriptsize
\begin{center}
\begin{tabular}{lc|cccc}
\hline
\hline
Epoch      & SI                    & $A$           & $B$            & $E$      & $F$ \\
\hline
1999-11-20 & $\alpha^{5.0}_{15.4}$ & -1.07\pl0.2   &   1.02\pl0.15  & $<-1.12$\pl0.12 & -              \\
2000-03-06 & $\alpha^{5.0}_{15.4}$ & -0.90\pl0.2   &   0.97\pl0.15  & $<-0.75$\pl0.15 & -              \\
2000-11-30 & $\alpha^{5.0}_{15.4}$ & -0.27\pl0.2   &   1.00\pl0.15  & $<-1.84$\pl0.08 & -              \\
2002-09-22 & $\alpha^{1.7}_{2.3}$  & $>7.80$\pl5.3 &   -            &  4.49\pl0.54    & $-2.49$\pl0.6  \\
           & $\alpha^{2.3}_{5.0}$  &  3.37\pl0.57  & $>5.22$\pl0.53 & $-1.26$\pl0.19  & $-0.06$\pl0.34 \\
           & $\alpha^{1.7}_{5.0}$  & $>4.61$\pl0.06& $>4.47$\pl0.05 &  0.35\pl0.18    & $-0.74$\pl0.24 \\
\hline
\end{tabular}
\caption[Spectral indices of NGC\,3079 components.]{Spectral indices of
NGC\,3079 components. Limits have been calculated using three times
the image rms.}
\label{tab:spectra}
\end{center}
\end{table*}
\renewcommand{\arraystretch}{1.0}

We give in Table~\ref{tab:spectra} all spectral indices calculated
from the results in Table~2, ignoring the 22\,GHz flux
densities and using the convention that
$S_\nu\propto\nu^\alpha$. Unfortunately, we do not have data at similar
frequencies before and after the deceleration events seen in
Figure~\ref{fig:sep+fluxes}, so we cannot analyse the spectral
evolution of components during these events.  Although our
observations were designed to avoid resolution effects, we cannot rule
out that the spectral index of $F$ is affected by its extent and by
some of the 2.3\,GHz flux density being missing due to the lack of
short baselines.

\section{Discussion}

We discuss the observational results as follows. In
Section~\ref{sec:si} we briefly comment on the spectral indices
observed in NGC\,3079; in Section~\ref{sec:jet-interact} we show that
components $A$ and $B$ are radio components in a jet which is
undergoing interaction with the surrounding material;
Section~\ref{sec:noboost} demonstrates that the behaviour of $A$ is
unlikely to be due to jet bending and Doppler boosting; and
Section~\ref{sec:E+F} shows that components $E$ and $F$ are
by-products of interactions of a jet with the surrounding material.

\subsection{Spectral indices}
\label{sec:si}

During the epochs of 1999 and 2000, the spectral indices
$\alpha^{5.0}_{15.4}$ of components $B$ and $E$ remained constant, and
component $A$ increased from -0.90 to -0.27 within the 8.5 months
prior to the observations in November 2000.

The spectral indices measured simultaneously in September 2002 are
more interesting. First, $A$ has a strongly inverted spectrum with
$\alpha^{2.3}_{5.0}=3.37$, as has $E$ with
$\alpha^{1.7}_{2.3}=4.49$. Second, $F$ has a very steep spectrum
between 1.7\,GHz and 2.3\,GHz, and there are several similar limits
which also deserve explanation.

The inverted spectra of $A$ and $E$ in the September 2002 epoch can be
modelled in terms of a free-free absorber in front of the components
(similarly inverted spectra and limits can be explained using the same
reasoning, but we lack the data for modelling). In the presence of a
free-free absorber, the observed flux density, $S_{\nu}$, of a
component can be related to the intrinsic flux density at 1\,GHz,
$S_0$, of the component by

\begin{equation}
S_{\nu} = S_{0} \left(\frac{\nu}{\rm GHz}\right)^{\alpha_0}\times exp(-\tau_{ff}),
\label{eq:flux}
\end{equation}

where $\alpha_0$ is the intrinsic spectral index and $\tau_{ff}$ is
the optical depth of the absorber with

\begin{equation}
\tau_{ff} = 0.0824\left(\frac{T}  {\rm K}      \right)^{-1.35}
                  \left(\frac{\nu}{\rm GHz}    \right)^{-2.1}
             \int       \frac{N_+}{\rm cm^{-3}}       
                        \frac{N_-}{\rm cm^{-3}}       
                        \frac{ds} {\rm pc}
\end{equation}

\begin{figure}
\centering
\includegraphics[width=\linewidth]{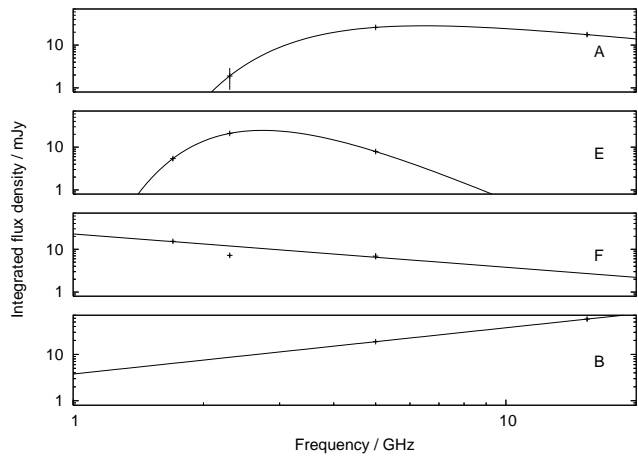}  
\caption{Spectra of components $A$, $E$, $F$, and $B$, measured in
September 2002 and using 15\,GHz data from November 2000. In the cases
of components $A$ and $E$ the lines represent the model described in
Section~\ref{sec:si}, whereas simple power-laws have been used for $B$
and $F$.}
\label{fig:spectra}
\end{figure}

(\citealt{Osterbrock1989}). The temperature, $T$, is commonly assumed
to be $10^4$\,K, $N_+$ and $N_-$ are the densities of positive and
negative charges in the absorber (assumed to be equal), and $ds$ is
the path length along the line of sight. The integral is called the
emission measure and is referred to as $k$
hereafter. Equation~\ref{eq:flux} has three unknowns, the intrinsic
flux density, $S_0$, the intrinsic spectral index, $\alpha_0$, and the
emission measure, $k$. It can be solved when the flux density has been
measured at a minimum of three frequencies. When upper and lower
limits are ignored, we can solve Equation~\ref{eq:flux} for $A$, $E$,
and $F$, taking into account the measurements at 1.7\,GHz, 2.3\,GHz,
5.0\,GHz, and 15.4\,GHz. In the case of $A$, we use the simultaneous
2.3\,GHz and 5.0\,GHz measurements from 2002 and add the 15.4\,GHz
measurements from November 2000. Hence variability may affect the
results. The models are plotted in Figure~\ref{fig:spectra}.

The spectrum of component F is concave and so is not described by a
free-free absorbed power law, and we have instead fitted a simple
power-law to the three spectral points. For component $A$ we find
$\alpha_0=-1.01$, $S_0=305\,{\rm mJy}$, and $k=7.4\times10^7\,{\rm
pc\,cm^{-6}}$, leading to an optical depth at 5\,GHz of 0.83. Thus $A$
appears to be a rather normal source of synchrotron emission behind a
moderately thick free-free absorber. In the case of $E$, however, the
intrinsic spectral index is $\alpha_0=-4.40$, the intrinsic flux
density is $S_0=17\,000\,{\rm mJy}$, and the emission measure is
$5.3\times10^7\,{\rm pc\,cm^{-6}}$, leading to $\tau_{ff}$ at 5\,GHz
of 0.59. Again the properties of the absorber are reasonable, but
$\alpha_0$ is difficult to explain.

One possible interpretation is that the population of synchrotron
electrons has been cut off from the central engine and has aged. The
population will lose the high-energy electrons first, leading to a
sharp drop in the observed spectral energy distribution towards higher
frequencies. The steep spectral index of $\alpha^{1.7}_{2.3}=-2.49$ of
$F$ could also arise from such a situation, for which we argue in
Section~\ref{sec:E+F}.

Another possibility is that the electron energy distribution is not a
power-law, but is a relativistic thermal distribution. Such models
have been considered by, e.g., \cite{Jones1979} and
\cite{Beckert1997}, and they usually reveal an exponential cutoff at
high frequencies. Both explanations are unsatisfactory because they
require the electron distribution cutoff to lie very close to the
frequency at which $\tau_{ff}=1$, but we can offer no better solution.

The result is robust against amplitude errors. Only when the
measurements at 1.7\,GHz and 5.0\,GHz are increased by as much as
50\,\%, and the 2.3\,GHz amplitude is decreased by 50\,\% does
Equation~\ref{eq:flux} yield ``common'' values for $\alpha_0$ of
component $E$. Amplitude errors of this magnitude, and in the required
sense, are unlikely.

Another possibility is resolution effects arising from the lack of
short interferometer spacings and gaps in the $(u,v)$-coverage, both
of which could lead to an underestimate of a partially resolved
emission component. In particular in the case of $F$ we cannot exclude
that the flux is underestimated at 2.3\,GHz or 5\,GHz. In this case
the spectral indices $\alpha^{1.6}_{2.3}$ and $\alpha^{2.3}_{5.0}$
given in Table~\ref{tab:spectra} should be regarded as lower
limits. For example, fitting a simple power-law spectrum for $F$ to
the three flux densities in Table~2 yields a spectral
index of $\alpha=-0.78\pm0.29$, which is close to the canonical value
of $\alpha=-0.7$.

\subsection{A jet interacting with the external medium at the location
of components $A$ and $B$}
\label{sec:jet-interact}


That components A and B are jet features is strongly supported by the
elongated structure between them (component $C$,
Fig.~\ref{fig:3079_C}, and component $D$, Fig.~3 in
\citealt{Trotter1998}), by their high speed of separation before 2003,
and because the line joining components $A$ and $B$ crosses the
dynamical centre of the accretion disk in NGC\,3079
(\citealt{Kondratko2005}). This position will be called the position
of the central engine hereafter.  The spectral properties of $A$ and
$B$ lead \cite{Trotter1998} (see also \citealt{Kondratko2005}) to
suggest that they are lower redshift and lower luminosity versions of
compact symmetric objects, whose observational properties are known to
be due to the interaction between jets and their high density external
medium.  These authors reported components $A$ and $B$ to have similar
spectra, which correspond to synchrotron self-absorbed or free-free
absorbed regions with turn-over frequencies in the range of 8\,GHz to
10\,GHz.  Moreover, \cite{Middelberg2005c} reported the spectra of the
radio continuum components within the innermost 1\,pc to 2\,pc from
the position of the central engine to be consistent with an scenario
in which the nucleus of NGC\,3079 is embedded in a dense medium.  All
these lines of evidence point to a jet-external medium interaction
scenario.

Our new results show that component $B$ has remained essentially
stationary both in position and in flux density since 2000.  In
contrast, component $A$ underwent two significant flux density
increases with two corresponding decelerations, the last one
completely halting the advance of this component.

Thus, these new results not only agree with the previously proposed
jet-feature nature of components $A$ and $B$ interacting with the
dense external medium, they also add further and stronger support to
this scenario.  The ``brighter-when-slower" behaviour of component $A$
is expected when a propagating jet encounters a progressively denser
external medium region (e.g. a non-homogeneous dense cloud).  The
numerical simulations of \cite{Wang2000} and \cite{Saxton2005} are
clear about what happens in that case.  If the jet is ``weak"
(i.e. much less denser and/or much less energetic) with respect to the
external dense cloud\footnote{see \cite{Wang2000} for the definition
of the parameters controlling the jet weakness with regard to its
external cloud, and for examples of weak and powerful jets}, the
impact slows down the advance of the ``hot-spot" of the jet and raises
its pressure.  This pressure increase enhances the synchrotron
emission.  In the approximation of \cite{Saxton2001,Saxton2002}, an
increase of the synchrotron emissivity\footnote{$j_{\nu}\propto
p^{(3-\alpha)/2}$, where $p$ is the pressure and $\alpha$ is the
spectral index of the synchrotron emission. We assume $\alpha\approx
-0.7$ here.} by a factor of 10, as experienced by component $A$
between 1997 and 2005 (see Fig.~\ref{fig:sep+fluxes}), would occur if
the pressure increased by a factor of $\sim$3.5.

\subsection{Unlikely jet Doppler boosting scenario for component $A$}
\label{sec:noboost}

Note that there is still a scenario that has not been proposed before
and that might, in principle, explain the ``brighter-when-slower"
behaviour of component $A$.  This scenario also involves a jet but the
reason for component $A$'s deceleration and brightening is the
change of propagation direction of component A towards the line of
sight, together with the Doppler boosting of the synchrotron radiation.

If the emission of component $A$ is beamed relativistically then the
observed flux density, $S_{\nu}$, and the intrinsic flux density,
$S'_{\nu'}$, are related by

\begin{equation}
S_{\nu} = \delta^3 S'_{\nu'}
\end{equation}

where $\delta$ is the Doppler factor\footnote{$\delta=[\Gamma(1-\beta
cos \theta)]^{-1}$ where $\Gamma=(1-\beta^2) ^{-1/2}$ is the Lorentz
factor, $\beta$ is the speed in units of $c$ and $\theta$ is the angle
between the direction of the flow and the line of sight.}
(e.g. \citealt{Hughes1991}).  An increase in the observed flux
density of component $A$ by a factor of two (as was observed between
the end of 2002 and the beginning of 2004) would imply an increase of
$\delta$ by a factor of $2^{1/3}$, which would also be the minimum
required value of the Doppler factor after the beginning of 2004.  As,
by that time the condition $\theta \approx 0^{\circ}$ is required
(because the motion of $A$ in the plane of the sky has stopped), the
intrinsic bulk plasma speed is constrained to be $\beta
\gtrsim 0.23\,c$.  An angle of the jet to the line of sight of
$\sim 90^{\circ}$ (i.e. the jets are in the plane of the sky) is
usually assumed if the accretion disk of the system is almost edge-on,
as in the case of NGC\,3079. The large acceleration required both to
bring the bulk plasma from $\beta
\approx 0.13\,c$ to $\beta \approx 0.23\,c$ and to bend the jet beam
by an amount of $45^{\circ}$ or larger, make the jet-Doppler boosting
scenario rather unlikely.    A lower angle ($\gtrsim 45^{\circ}$)
is assumed here to account for the fact that the jet is clearly not
aligned with the angular momentum of the disk (see e.g.
\citealt{Kondratko2005}) but at $\sim 45^{\circ}$ from such direction.

Hence, we conclude that the Doppler boosting scenario requires the jet
in NGC\,3079 at the location of $A$ to bend by $\sim45^{\circ}$ in a
time range of only one year. In such conditions, a strong impact with
the external medium unavoidably produces such extreme bending.  In
this case, the Doppler boosting scenario brings us again to the
interaction of the jet with the external medium.

\subsection{Components $E$ and $F$ as by-products of jet-external
medium interaction}
\label{sec:E+F}

%

The fact that components $E$ and $F$ were found at positions not
aligned with the jet between components $A$ and $B$ made
\cite{Kondratko2005} suggest that $E$ and $F$ were intrinsically
different from $A$ and $B$.  In addition, the spectral properties of
$E$ and $F$ (with much steeper spectra than those corresponding to
optically thin synchrotron spectra) led \cite{Kondratko2005} to
propose that $E$ and $F$ are consistent with ageing synchrotron
sources; i.e. with no injection of fresh non-thermal electrons.

\cite{Kondratko2005} showed that the possibility that $E$ and $F$
are radio supernovae is highly unlikely.  Instead, they proposed two
different scenarios to explain the properties of $E$ and $F$: {\it a)}
either they are remains of emission from a wobbling jet which was
ejected earlier in orientations different from the current jet
direction or {\it b)} they are shocks in a wide-angle, pc-scale outflow
interacting with its external medium.  Within the latter picture, the
outflow would be a pc-scaled version of the kpc-scale bipolar
super-bubbles observed in radio continuum emission
(\citealt{Duric1988,Baan1995,Irwin2003}), in [N II]+H$\alpha$ emission
(\citealt{Veilleux1994,Cecil2001}) and in soft X-ray continuum
emission (\citealt{Cecil2002}).

We consider the jet wobbling scenario ({\it a}) very unlikely since no
evidence has been found to support a systematic change of the position
angle of the current jet. In this scenario, the jet axis defined as
the line which connects components $A$ and $B$, should systematically
change with time, which has not been observed between 1994 and 2005
(cf. Table~\ref{tab:pos}). Furthermore, component $C$ nicely lines up
with the $A$-$B$ axis in all images since \cite{Irwin1988} observed
NGC\,3079 for the first time with VLBI. Hence, during this time range,
the jet wobbling scenario can be safely ruled out.

Although scenario ({\it b}) is not as unlikely as the jet wobbling
scenario, we still find it difficult to imagine an outflow ejected
from the central engine in NGC\,3079 which interacts with the dense
external medium and produces a shock moving \emph{inwards}, in the
opposite sense to the assumed \emph{outflow}, as detected for
component $E$ in 1999-2000 (Figure~\ref{fig:xy}).

Here, we propose an alternative scenario to account for the presence
of synchrotron emission regions from non-thermal aging electrons close
to the pc-scale jets in NGC\,3079.  This new scenario comes up
naturally from the jet/dense-external-medium interaction picture
previously proposed to explain the behaviour of component A.  Our idea
is based on the two-dimensional numerical simulations of jets in
inhomogeneous media reported by \cite{Saxton2005}\footnote{The scaling
of the results from \cite{Saxton2005} was set to apply their
quantitative results to kpc-scale jets.  However, for us who use their
study on a qualitative way only, their results are equally
applicable.}.

Their grid of simulations, on which a tenuous medium is also populated
by a relatively small filling factor of dense clouds, show that when
the filling factor of the clouds is large, the jet has more
possibilities to be disrupted and to form channels through which the
shocked radio emitting plasma can flow between the clouds.  The
formation of a hierarchical structure of nested lobes (not always
producing strong radio emission) and an outer enclosing bubble seems
unavoidable in all the simulated conditions.  Having in mind this
scenario, it is difficult not to find a close, at least qualitative,
relation between our component $A$ with the region of the simulated
jet in which the first dense cloud is encountered and the jet is
disrupted; and between components $E$ and $F$ with the simulated lobes
or knots of already shocked material.  In the simulations, this
material slowly flows from the first shocked region along the opened
channels between the clouds (see e.g. Fig.\,11 of
\citealt{Saxton2005}) and not necessarily outwards, away from the
central engine, because their motions depend on the cloud-obstacles
that they encounter in propagation. This could explain why component
$E$ does not move outwards but inwards, towards the central engine
(Figure~\ref{fig:xy}).  The simulations of \cite{Saxton2005} show that
the lobes of shocked material can be found at large distances
(relative to the jet size) from the first shocked region and in
directions that are far from where the original straight jet would
have propagated.

A consequence of this is that, as the clouds move, plasma channels
that initially emit synchrotron emission can eventually be closed,
cutting off synchrotron-emitting regions from a fresh supply of
non-thermal particles. As a consequence, the electron population of
synchrotron-emitting regions is expected to age. This process
manifests itself in an overall decrease of flux density and the
steepening of radio spectra, as high-energy particles loose their
energy more quickly and radio emission at higher frequencies decreases
first. In NGC\,3079, the 5\,GHz flux density of component $E$ has
approximately doubled between 1999 and 2005, which seems to disagree
with the scenario of an aging electron population. However, we note
that $E$ underwent significant internal evolution in 1999 and
2000. This evolution maybe caused by compression of the component,
which caused the observed increase in flux density.

\section{Summary and Conclusions}

Based on the similarities of the spectra of GPS sources with those of
components $A$ and $B$, previous studies proposed these components to
be regions in which a jet, launched from the central super-massive
engine in NGC\,3079, interacts with the dense external medium.
Moreover, \cite{Middelberg2005c} reported the spectra of the radio
continuum components within the innermost 1\,pc to 2\,pc from the
accretion disk to be consistent with an scenario in which the nucleus
of NGC\,3079 is embedded in a dense medium.  We have shown that the
structure and the time evolution of the radio continuum emission in
the innermost pc around the accretion disk in NGC\,3079 behaves as
expected for a jet interacting with a clumpy external medium. Hence,
our new results not only agree with both the jet scenario and the
jet-interaction scenario previously proposed, but also strongly
support both of them.

Within our proposed scenario, depending on the properties of the
cloudy medium in which they are embedded, $A$ and $B$ are expected to
continue separating from the dynamic centre of the accretion disk once
they are able to drill through the cloud that stopped them, as was
proposed for the expanding phase of the jet in III\,Zw\,2
(\citealt{Brunthaler2005}).  Note however, that the re-activation of
the expansion of components $A$ and $B$ might start to be significant
after large time periods (depending on the relative properties of the
jets and the clumpy medium on which they are embedded). Thus, these
expansions may not be observable for years or tens of years.  The
spectra of components $E$ and $F$ should show evidence of a slowly
ageing non-thermal electron population until they have faded
completely.  Although not required, they can be expected to move with
speeds smaller than those previously measured for $A$ and in
directions which do not necessarily have to point outwards from the
central engine in NGC\,3079.

The simulations of \cite{Saxton2005} show that, if the filling factor
of dense clouds in the surrounding medium of the jet is high, the
momentum of the jet spreads at the first strong interaction and forms
a slowly expanding bubble which grows larger than the jet itself,
which indicates that the jet is impeded from propagating freely.  If
one assumes a large filling factor of dense clouds surrounding the
central region of NGC\,3079, the interaction scenario offers, as a
natural consequence of the slower propagation of the jet, the
explanation for the origin of the kpc-scale bipolar super-bubbles
observed in radio continuum emission, [N II]+H$\alpha$ emission and
soft X-ray continuum emission in NGC\,3079.

There is at present little doubt about the presence of strong
interactions of the radio plasma with the inter-stellar medium in
Seyfert galaxies (\citealt{Oosterloo2000}).  As for NGC\,3079,
evidence for interaction of the jets with their external medium in
subgalactic-scales seems to be common in Seyfert galaxies
(e.g. NGC\,1068 and NGC\,4151, \citealt{Wilson1982}; III\,Zw\,2,
\citealt{Brunthaler2005}). Hence, it appears reasonable to apply the
interaction scenario to the whole class of Seyfert galaxies.  This
would give a satisfactory explanation for why in a large fraction of
Seyfert galaxies large-scale bipolar super-bubbles are observed
(e.g. \citealt{Colbert1996a}, \citealt{Colbert1996b}) instead of well
collimated, kpc-scale jets as in radio-loud AGN.

\section*{Acknowledgments}

I. Agudo acknowledges financial support from the EU Commission for
Science and Research through the ENIGMA network (contract
HPRN-CT-2002-00321). The VLBA is an instrument of the National Radio
Astronomy Observatory, which is a facility of the National Science
Foundation of the U.S.A. operated under cooperative agreement by
Associated Universities, Inc. (U.S.A.). We thank M. Perucho for
reviewing the manuscript.

\bibliography{refs}

\label{lastpage}

\end{document}